\documentclass[aps,groupedaddress,nofootinbib,11pt]{revtex4}

\usepackage[dvips]{epsfig}
\usepackage{amsfonts}
\usepackage{amsmath}
\usepackage{}

\newcommand{\Eqref}[1]{(\ref{#1})}
\newcommand{\ket}[1] {\mbox{$ \vert #1 \rangle $}}
\newcommand{\kett}[1] {\mbox{$ \vert #1 \rangle_2 $}}
\newcommand{\bra}[1] {\mbox{$ \langle #1 \vert $}}
\newcommand{\abs}[1] {\mbox{$ \vert #1 \vert $}}

\newcommand{\inv}[1]{\frac{1}{#1}}

\newcommand{\av}[1]{\langle\!\langle \,  #1 \, \rangle\!\rangle}

\newcounter{subequation}[equation] \makeatletter
\expandafter\let\expandafter\reset@font\csname reset@font\endcsname
\newenvironment{subeqnarray}
  {\arraycolsep1pt
    \def\@eqnnum\stepcounter##1{\stepcounter{subequation}{\reset@font\rm
      (\theequation\alph{subequation})}}\eqnarray}
  {\endeqnarray\stepcounter{equation}}
\makeatother

\newcommand{\ba}{\begin{eqnarray}}
\newcommand{\ea}{\end{eqnarray}}
\newcommand{\sba}{\begin{subeqnarray}}
\newcommand{\sea}{\end{subeqnarray}}

\def\th{\mbox{th}}

\def\sh{\mbox{sh}}

\begin{document}

\vskip 1truecm

\title{
Inflationary 
spectra, decoherence, 
and two-mode coherent states}
\vskip 1truecm
\author{David Campo}
\email[]{campo@phys.univ-tours.fr}
\affiliation{Laboratoire de Math\'{e}matiques et Physique
Th\'{e}orique, CNRS UMR 6083,
Universit\'{e} de Tours, 37200 Tours, France}
\author{Renaud Parentani}
\email[]{Renaud.Parentani@th.u-psud.fr}
\affiliation{Laboratoire de Physique Th\'{e}orique, CNRS UMR 8627,
B\^atiment 210,
Universit\'{e} Paris XI, 91405 Orsay Cedex, France}
\maketitle

\vskip 1truecm
\centerline{{\bf Abstract }}
\vskip 0.5truecm

We re-examine the question of the entropy stored in the 
distribution of primordial density fluctuations. To this end
we make use of two-mode coherent states. These states
incorporate the isotropy of the distribution
as well as the temporal coherence and the
semi-classical character of highly amplified modes. 
They also provide a lower bound for the 
entropy 
if, as one expects, decoherence processes erase the squeezing
which originally characterized the distribution in inflationary models.
This lower bound is one half the maximal (thermal) value.
By considering backreaction effects, we also provide an upper bound
for this entropy at the onset of the adiabatic era.

\vskip 1truecm 

\section{Introduction}

Inflation tells us that the primordial
density fluctuations arise from the amplification 
of vacuum fluctuations \cite{Mukhanov81,Staro79}. As a result
of this amplification,
the initial vacuum state becomes a product of 
highly squeezed two-mode states \cite{Grishchuk90}.
In spite of the complexity of this 
state, 
when computing expectation values, i.e. the two-point function,
the modes exhibit a temporal coherence upon horizon re-entry.
When considering the physics which took place near the recombination, 
it is therefore convenient and sufficient to enforce the 
temporal coherence by 
putting to zero
the decaying mode.  
Then the residual random properties consist in 
treating the amplitude of the growing mode as a stochastic
variable, thereby ignoring the quantum properties
of the original Gaussian distribution.

However this simplified description has several drawbacks.
In particular the settings are too restrictive
to describe the distribution which 
would result from some decoherence process
which would have taken place during the early universe.
More generally, the simplified settings are unable
to 
parameterize 
 deviations from the standard results
which preserve
the isotropy and the Gaussianity of the distribution.

In this paper, we show that the appropriate basis to investigate 
these questions is provided by two-mode coherent states. 
The reasons are the following. 
First, a two-mode coherent state provides the quantum counterpart
of a particular classical realization of the ensemble
of metric fluctuations. This correspondence is well defined
for the highly excited modes we are dealing with. 
(Remember that the observed temperature anisotropies of relative amplitude 
$10^{-5}$ fix the occupation number $n$
to be of the order of $10^{100}$.)  
Second, ensembles of two-mode coherent states
can characterize 
{\it any} Gaussian isotropic distribution,
thereby allowing to describe 
arbitrary levels of coherence. 
This follows from the requirement of isotropy and homogeneity
which
restricts the non-vanishing matrix elements
of the distribution: only two-mode elements with opposite wave vectors
could be different from zero.
Third, they allow to make contact with the general remark
of \cite{Zurekbook,Zurek93}
according to which squeezed states rapidly decay
into a statistical mixture of coherent states when small non-linearities
are no longer neglected\cite{Maldacena03}. 
In fact when applying this general theorem to 
highly squeezed two-mode states,
the resulting 
distribution is precisely a 
diagonal density matrix of two-mode coherent states. 

This distribution defines the {\it minimal} entropy 
stored in the primordial spectrum
(or in other words, the minimal coarse graining) 
when no squeezing remains,
that is, when there is no 
direction in phase space in which the spread 
of the non-diagonal elements of the density matrix 
is 
smaller than that of the vacuum.
For each 
mode the entropy bound is $1/2$ the maximal (thermal) value, 
given the occupation number.
We shall see that
this entropy coincides with that associated  with
the simplified prescription which consists in neglecting the 
decaying mode. It is to be stressed that this identification can be
reached because in the large occupation number limit, the entropy
is extremely sensitive to the level of decoherence whereas the power spectrum is 
instead extremely robust. Typically the relative modifications 
of the latter are $1/n$ whereas the changes of the former are in $\ln n$.
  
The open question concerns the efficiency of 
decoherence processes in nature: 
are they powerful enough to suppress
the squeezing that the initial density matrix possessed?
This question is currently under investigation. 
Preliminary results indicate that the squeezing is indeed erased,
thereby implying that the resulting entropy is larger than (or equal to)
the above mentionned bound.


Finally, we shall also provide an 
upper bound for this entropy by considering
backreaction effects at the end of the inflationary period.
This upper bound is given by $3/4$ of the maximal entropy.
The evaluation of these entropies is exactly performed
by exploiting the fact that any
Gaussian isotropic distribution can be expressed in terms of
thermal distributions, see Appendix C. 

Related questions have been already analyzed
in several papers, see 
\cite{MukhaBrandyProkopecg,Albrecht94,Giovanninig,Matacz94,PolStar96,KiefPolStar98a,KiefPolStar00}. 
What we add in the present paper is 
a further clarification of the matters,
the usefulness and the relevance of two-mode coherent states,
 the lower and upper bounds on the entropy, and relationships
between various elements which have been some how separately discussed.
Notice that 
we shall not discuss the physical relevance of this entropy for structure
formation. For this interesting question we refer to \cite{MukhaBrandyProkopecg}.



\section{A review of the standard derivation of primordial spectra}

\subsection{Quantum distribution of two-mode states}

In this subsection we recall how the amplification of modes 
of a massless field propagating in a FRW universe translates 
in quantum settings in the fact that 
the initial ground state
evolves into a product of
highly squeezed two-mode states. 
Before proceeding, we remind the reader that 
it has been shown that the evolution of linearized cosmological perturbations
(metric and density perturbations) reduces to the
propagation of 
massless
scalar fields in 
a FRW spacetime \cite{MBF92}.
In this article, we shall only consider the massless scalar test field
since the transposition
of the results 
to physical fields represents no difficulty.
Indeed, when preserving
the linearity of the evolution,
the only modification concerns the late time dependence of the modes.

Let us work in a flat FRW universe. 
The line element is:
\ba \label{metric}
 ds^2 &=& a(\eta)^2 \left[ -d\eta^2 +  \delta_{ij}dx^{i}dx^{j} \right] \, .
\ea
For definiteness and simplicity, we consider a 
cosmological evolution which starts with an inflationary de Sitter
phase and ends
with a 
radiation dominated period.
When using the conformal time $\eta$ to parametrize the
evolution, the scale factor is respectively given by 
\sba \label{aeta}
 a(\eta) &=& - \frac{1}{H(\eta-2 \eta_r)} \, , \qquad 
 {\mbox{for\, \, }} -\infty < \eta < \eta_r  \, , \\
 a(\eta) &=& \frac{1}{H\eta_r^2} \, \eta  \, , \qquad 
  {\mbox{for\, \, }}  \eta > \eta_r \, , 
\sea
where $\eta_r> 0 $ 
designates  
the end of inflation. 
The transition
is such that
the scale factor and the Hubble parameter are
continuous functions.
This approximation based on an instantaneous transition is 
perfectly justified for modes 
relevant for CMB physics. Indeed, 
their wave vector $k$ 
obeys 
$k \eta_r  \sim 10^{-25}\sim e^{-60}$ when inflation
lasts for $60$ e-folds.  
Hence, the 
phase shift they could accumulate in a more realistic smoothed out transition
is completely negligible.

Let $\xi(\eta,{\bf x})$ be a massless test scalar field 
propagating in this background metric.
It is convenient to work with the rescaled field $\phi = a \xi $ and
to decompose it into Fourier modes
\ba
\phi(\eta, \bold x)
&=& \int\!\!d^3k \, \frac{e^{i \bold k \bold x}}{(2\pi)^{3/2}} 
  \phi_{\bf k}(\eta) \, .
\ea
The time dependent mode $\phi_{\bold k}$ obeys 
\ba \label{PropPhi_k}
 \partial_{\eta}^2 \phi_{\bf k}  + 
 \left(k^2 - \frac{\partial_\eta^2 a}{a} \right) \phi_{\bf k} = 0 \, ,
\ea
where $k=\abs{\bold k}$ is the norm of the conformal wave vector. 
  
In our background solution,
${k}^2 - {\partial_\eta^2 a}/{a}$ 
is negative during the de Sitter period when the wavelength
is larger that the Hubble radius. 
This leads to a large amplification of $\phi_k$. 
In quantum terms this mode amplification translates 
into spontaneous pair production characterized by 
correspondingly large occupation numbers.

To obtain the final distribution of particles,
one should introduce two sets of 
positive frequency solutions of Eq. \Eqref{PropPhi_k}. 
The in modes are defined at 
asymptotic early time, and the out ones at late time. Both have 
unit positive Wronskian in conformity with the usual
particle interpretation \cite{B&D}. One gets  
\sba \label{modes}
 \phi_k^{in}(\eta) &=& \frac{1}{\sqrt{2k}} 
 	\left(1-\frac{i}{k(\eta - 2 \eta_r)}\right)
	e^{-ik(\eta - 2 \eta_r)} \, , \quad 
 \mbox{for\, \,} \eta < \eta_r   \, , \\
 \phi_k^{out}(\eta) &=& 
 	\frac{1}{\sqrt{2k}} e^{-ik\eta} \, ,\quad 
  \mbox{for\, \, }  \eta > \eta_r  \, .
\sea
In spite of the time dependence of the background,
these two positive frequency modes are unambiguously 
defined (up to an arbitrary constant
phase which drops in all expectation values and which has here been chosen
so as to simplify the forthcoming expressions).
In the radiation dominated era there is no ambiguity 
since the conformal frequency is constant because $\partial^2_\eta a = 0$. 
In the de Sitter epoch, there is no ambiguity either for relevant modes 
if inflation lasts more than $70$ e-folds, see \cite{NPC02} for the evaluation
of the small corrections one obtains when imposing vacuum at some finite 
early time. 
Similarly, in quantum settings, there is no ambiguity
for the initial state of relevant modes: at the onset of inflation
they must be in their ground state\cite{MBF92,cmbParenta}.

The in and out modes are related by a Bogoliubov transformation 
\ba \label{inoutmodes}
 \phi_k^{in}(\eta) = \alpha_k \phi_k^{out}(\eta) + 
 		     \beta_k^* \phi_k^{out\, *}(\eta) \, ,
\ea
where the Bogoliubov coefficients are given by the Wronskians
\ba \label{abcoef}
  \alpha_{k} = \left( \phi_{k}^{out} , \phi_{k}^{in} \right)
  \mbox{, \ \ }
  \beta_{k}^{*} = -  \left( \phi_{k}^{out\, *} , \phi_{k}^{in} \right) \, .
\ea 
These overlaps should be evaluated at transition time $\eta_r$ 
since modes satisfy different equations in each era.
One gets
\sba \label{bogolincosmo}
 \alpha_k &=& - \frac{ e^{2i k \eta_r}}{2k^2\eta_r^2} 
 \left( 1 - 2 i{k\eta_r}- 2k^2\eta_r^2 \right) 
 = \frac{-1}{2k^2\eta_r^2} \, ( 1 + O(k\eta_r)^3) \, , \\
 \beta_k^{*} &=& 
 \frac{1}{2k^2\eta_r^2} \, .
\sea
Thus, for relevant modes, $k\eta_r \sim 10^{-25}$,
the in modes are enormously amplified. Concomitantly, they are 
dominated by the sine during the radiation dominated era:
\ba \label{neglectdecay}
 \phi_k^{in}(\eta) 
 &=& \frac{i}{k^{2}\eta_r^2} \left[ \frac{\sin k\eta}{\sqrt{2 k}} \, 
      + \, O(k\eta_r)^3 \,  \cos k\eta   \right]
      \, , \qquad  \eta \geq \eta_r \, .
\ea
Once the cosine in Eq. \Eqref{neglectdecay} is neglected, the physical modes 
$\phi^{in}_k/a$ show the same temporal behaviour, e.g. they are constant until
they start oscillating as they re-enter the Hubble radius when $k \eta \simeq 1$.

Lets now see how these considerations translate in second quantized settings.
Each mode operator is decomposed twice 
\ba \label{hatphik}
 \hat \phi_{\bold k}(\eta) = 
 \hat a_{\bold k}^j  \phi_{k}^j(\eta) + 
 \hat a_{-\bold k}^{j\, \dagger} \phi_{k}^{j \, *}(\eta) \, ,
\ea
where $j$ stands for both the in and out basis.
The operators so defined are related by the 
transformation 
\ba \label{inoutoperator}
 \hat a_{\bold k}^{in} = \alpha_k^{*} \, \hat a_{\bold k}^{out} -
		      \beta_k \, \hat a_{-\bold k}^{out \, \dagger} \, .
\ea 
This transformation 
couples ${\bf k}$ to $-{\bf k}$ only.
Hence, when starting from the in vacuum (the state annihilated by the
$\hat a_{\bold k}^{in}$ operators), 
every out particle 
of momentum ${\bf k}$ will be accompanied by a partner 
of momentum $-{\bf k}$. Moreover, pairs characterized 
by different momenta are incoherent 
(in the sense that in expectation values any
product of annihilation and creation operators of different momenta 
will factorize).

These two properties are explicit when  
writing the in vacuum in terms of out states (i.e.
states with a definite out particle content). From Eq. \Eqref{inoutoperator}, 
one gets (see \cite{MukhaBrandyProkopecg,PhysRep95})
\ba \label{inoutvac}
 \ket{0,\, in} 
&=&  
 \widetilde{\prod\limits_{\bold k}} \otimes  
 \kett{0,\, \bold k, in} 
\nonumber\\ 
&=& 
\widetilde{\prod_{\bold k}}  \otimes 
\left(  \inv{\vert \alpha_{k} \vert} 
   \exp \left( z_{k}\, 
  \hat  a_{\bold k}^{out \, \dagger}  \, 
  \hat a_{-\bold k}^{out \, \dagger} \right)
  \ket{0,\, \bold k, out}\otimes \ket{0,\, -\bold k, out} \right) \, .
\ea
The tilde tensorial product takes into account only half the modes.
It must be introduced 
because the squeezing operator acts both on the $\bf k$ and the $-\bf k$
sectors. 
The definition of this product 
requires the introduction of an arbitrary wave vector 
to divide the modes into two sets. The sign of $k_z$ can be used. 
Notice that a rigorous definition of $\widetilde{\prod_{\bold k}}$
requires to consider a
discrete set of modes normalized with Kronekers (that is, to normalize the 
modes in a finite conformal 3-volume).
To be explicit, the two-mode state $\kett{0,\, \bold k, in} $
is given by
\ba
\label{in2mode}
 \kett{0,\, \bold k, in} = 
 \ket{0,\, \bold k, in} \otimes \ket{0,\, -\bold k, in} \, ,
\ea 
where $\ket{0,\, \bold k, in}$ is the ground state of the 
$\bold k$-th mode at the onset of inflation.
The complex parameter 
$z_k$ appearing in the squeezing operator
in Eq. \Eqref{inoutvac}
is given by the ratio of the Bogoliubov coefficients
\ba \label{z}
 z_k &=& \frac{\beta_k}{\alpha_k^*} = - e^{-2i k \eta_r} 
 \frac{1}{\left( 1 + i2{k\eta_r}- 2k^2\eta_r^2 \right)} 
 \nonumber \\
 &=& -  1 +  O(k\eta_r)^3  \, .
\ea
The high occupation number limit corresponds to $\abs{z_k} \to 1^-$.

It has to be emphasized that none of the out states in Eq. \Eqref{inoutvac} 
carry
3-momentum. Hence, the distribution is homogeneous in a strong sense:
at late time the 3-momentum operator is still
annihilated by the state of Eq. \Eqref{inoutvac}. 
(This property is not satisfied 
by incoherent distributions such as thermal baths. In those cases, 
the 3-momentum fluctuates and vanishes only in the mean.)
The present distribution is also isotropic
since the Bogoliubov coefficients are functions of the norm $k$ only. 
Finally, it is a Gaussian distribution, as can be seen from 
Eq. \Eqref{inoutvac}. 

To appreciate the peculiar properties of the distribution 
of Eq. \Eqref{inoutvac} it is interesting to consider the most 
general
homogeneous, isotropic, and Gaussian distribution of out quanta.
Its properties are completely specified by three real functions of the norm
$k$ (one real and one complex) through the following expectation values 
\sba \label{momenta}  
 && \langle \hat a_{\bold k}^{out} \rangle
 =  0 \, , \\
 &&\langle \hat a_{\bold k}^{out\, \dagger} \,  
 \hat  a_{\bold k'}^{out} \rangle
 = n_k   \, \delta^3(\bold k - \bold k' ) \, ,
\\
 &&\langle \hat a_{\bold k}^{out}  \,  
 \hat a_{\bold k'}^{out} \rangle
 = c_k \, \delta^3(\bold k + \bold k' )\, .
\sea 
In the second line, $n_k$ is the mean occupation number.
In the third one, the complex number $c_k$ 
characterizes the quantum coherence of the distribution.
The degree of two-mode coherence is given by $|c_k|/(n_k+1/2)$, see Appendix C.
It is bounded by 1.
For a thermal (incoherent) distribution, one has $c_k \equiv 0$: no coherence.

In the  case of pair production from vacuum, 
one has 
\ba
\label{cterm}
 n_k &=& \abs{\beta_k}^2 = \frac{\abs{z_k}^2}{1- \abs{z_k}^2}\, ,
 \nonumber\\
 c_k 
 &=& \alpha_k \beta_k = 
 \frac{z_k}{1- \abs{z_k}^2}
 \, .
\ea
Therefore, when considering relevant modes in our inflationary model 
Eq. \Eqref{bogolincosmo},
one has 
\ba
\label{cterm2}
 n_k &=& \frac{1}{4 (k \eta_r)^4 } \simeq 10^{100}\, ,
 \nonumber\\
\frac{\vert c_k \vert}{ n_k+1/2} &=& 1 + O(k \eta_r)^3\, . 
\ea
That is, the distribution 
which results from inflation is highly populated and, more importantly,
maximally coherent. 
When computing the Green function,
the two-mode coherence of the distribution
will manifest itself in the {\it temporal coherence} of the modes.

\subsection{Two-point function and the neglect of the decaying mode}

When expressed in terms of out modes,
the two-point function associated with the general 
distribution specified by Eqs. \Eqref{momenta} is
\ba \label{Ginx}
 G(\eta,{\bf x}, \eta', {\bf x'}) &=& G_{out}(\eta,{\bf x}, \eta', {\bf x'})  
 \nonumber \\
 &&
+ \int_0^{\infty} \frac{dk k^2}{\pi^2} 
 \frac{\sin(k\vert {\bf x} - {\bf x}'\vert)}{k\vert {\bf x} - {\bf x}'\vert} 
 2 Re \left[  n_k \, \phi_k(\eta) \phi_k^*(\eta')  
+  c_k \, \phi_k(\eta) \phi_k(\eta') 
\right]  . \quad
\ea
In the first line, we have isolated $G_{out}$,
the contribution of the out vacuum. In the high occupation 
number limit, this contribution is negligible.

It is important to notice that, for a general distribution, 
the sum in bracket in the
above integrand 
cannot be factorized. 
However it does factorize 
in two cases which are relevant for us: first,
 for coherent states, see Appendix A, and second 
for distributions resulting 
from pair production from vacuum.
Indeed, when considering our cosmological model, 
taking into account the minus sign of Eq. \Eqref{z},
and neglecting correction terms in $O(k\eta_r)^3$
(which amounts to neglect the decaying mode, see Eq. \Eqref{neglectdecay}),
one obtains
\ba \label{Ginx2}
G_{in}(\eta,{\bf x}, \eta', {\bf x'})
= \int_0^{\infty} \frac{dk k^2}{\pi^2} 
 \frac{\sin(k\vert {\bf x} - {\bf x}'\vert)}{k\vert {\bf x} - {\bf x}'\vert} 
 \,   n_k \,  \frac{\sin k\eta}{\sqrt{k}}\, \frac{\sin k\eta'}{\sqrt{k}}
 \, .
\ea
As announced, the integrand in Eq. \Eqref{Ginx} 
factorizes. In inflation, the function which appears is $ \sin k\eta$
where $\eta$ is related to the scale factor by Eq. \Eqref{aeta}.
This is how the temporal coherence of modes obtains
from the two-mode coherence of the distribution.

Once the cosine is neglected,
the quantum distribution 
is effectively replaced by 
a stochastic Gaussian distribution of classical fluctuations 
\ba 
\label{sineeq}
 \phi_{\bf k}(\eta) = S_{\bf k}\frac{\sin k\eta}{\sqrt{k}}  \, ,
\ea
with locked temporal
argument, and random amplitudes with variances given by
\ba \label{Aeff}
 \av{S_{\bf k}S_{\bf k'}^{*}}_{e\!f\!f} =  \av{S_{\bf k}S_{-\bf k'}}_{e\!f\!f} = 
 2n_k \, \delta^{(3)}({\bf k}-{\bf k}') \, .
\ea
Being Gaussian,
the effective probability distribution is simply
\ba \label{Peff}
 {\cal P}_{e\!f\!f} = \widetilde{\prod_{\bold k}} \,
 \frac{1}{2n_k} \exp\left(-\frac{\abs{S_{\bf k}}^2}{2n_k} \right) \, .
\ea
To avoid double counting, one must again
 use the tilde product which takes into account half the modes
only, as was done in the quantum distribution of Eq. \Eqref{inoutvac}. 
This counting becomes crucial when computing the entropy, see Section III.C.

Notice finally that the first equality in Eq. \Eqref{Aeff}
is simply the expression of the reality of the field 
$\phi(\eta,{\bf x})$. This was not the case when noticing that inflation
gives
$\langle \hat a_{\bf k}^{out\, \dagger}  
\, \hat a_{\bf k'}^{out} \rangle = - 
\langle \hat a_{\bf k}^{out}  \, \hat a_{-\bf k'}^{out} \rangle
(1 + O(k \eta_r)^3)$ for the quantum field operators. 
In that case the equality is the expression of the coherence
of the in vacuum, i.e. the absence of 3-momentum fluctuations
and the possibility of factorizing the
2-point function by discarding the cosines.

\subsection{Additional remarks}

First
we 
remind the reader why a random 
distribution of both the sine and the cosine 
does not give rise to any temporal 
coherence \cite{Grishchuk90}. 
In fact such a distribution  corresponds to
an incoherent (thermal) distribution.

Writing the field 
modes as 
the sum of a sine and a cosine:
\ba
 \hat \phi_{\bf k} &=& \hat a_{\bf k}^{out} \phi_k^{out} + 
 \hat a_{-\bf k}^{\dagger \, out} \phi_k^{out \, *} \, \nonumber \\
 &=& \hat C_{\bf k} \frac{\cos(k\eta)}{\sqrt{k}} + 
 \hat S_{\bf k} \frac{\sin(k\eta)}{\sqrt{k}} \, .
\ea
It is to be noticed that 
the operators $\hat C_{\bf k}$ and $\hat S_{\bf k}$ 
are 
proportional to the field mode operator and its conjugate momentum evaluated
at $\eta = 0$, as if there were no inflation:  
\ba
 \hat C_{\bf k} &=&  
 \frac{1}{\sqrt{2}}  \, 
 \left(\hat a_{\bf k}^{out} + \hat a_{-\bf k}^{\dagger \, out} \right)
 = \sqrt{k} \,  \hat \phi_{\bf k}(0) 
 \, ,  \nonumber \\
 \hat S_{\bf k} &=& \frac{-i}{\sqrt 2}  \, 
 \left( \hat a_{\bf k}^{out} - \hat a_{-\bf k}^{\dagger \, out}\right) 
 = \frac{1}{\sqrt k} \, \partial_{\eta} \hat \phi_{\bf k}(0) 
 \, .
\ea
They
satisfy the canonical equal time commutation relations.

Consider an incoherent ($c_k =0$) distribution:
\ba
\label{incoh}
\langle \hat a_{\bf k}^{\dagger} \hat a_{{\bf k}'} \rangle_{i\!n\!c}
= n_k \, \delta^{3}({\bf k}-{\bf k}') \, ,\qquad
\langle \hat a_{\bf k} \hat a_{{\bf k}'} \rangle_{i\!n\!c} = 0 \, .
\ea
Then $\hat C_{\bf k}$ and $\hat S_{\bf k}$ are
uncorrelated Gaussian operators with equal variance:
\ba
 \langle {\hat  C_{\bf k}\hat C_{{\bf k}'}^\dagger}\rangle_{i\!n\!c} &=& 
\langle  {\hat S_{\bf k}\hat S_{{\bf k}'}^\dagger}\rangle_{i\!n\!c} 
= \left( n_k + \frac{1}{2} \right) \, \delta^{3}({\bf k}-{\bf k}') 
 \, , \nonumber \\
\langle  {\hat C_{\bf k}\hat S_{{\bf k}'}}\rangle_{i\!n\!c} &=& 0 \, .
\ea
This implies the absence of temporal coherence of the modes,
as can be seen from the temporal behaviour of the bracket in the 
integrand of Eq. \Eqref{Ginx}: when $\eta = \eta'$  the bracket does not 
exhibit any oscillation in $k$ as it did for the inflationary distribution. 
(This absence can also be understood by
considering the $S_{\bf k}$ and $C_{\bf k}$ 
as stochastic variables rather than quantum ones.
Writing 
the mode in terms of its norm and its phase \cite{Grishchuk90,Allen00}
\ba 
 \phi_{\bf k} &=& \Phi_{\bf k} \, \sin(k\eta + \theta_{\bf k}) \, .
\ea
one can treat $\Phi_{\bf k}$ and $\theta_{\bf k}$ as stochastic variables.
One then verifies that the distribution for the phase $\theta_{\bf k}$ is uniform
over the interval $\left[0,\, 2\pi \right]$. Hence no temporal coherence 
could obtain.)


As a last remark, to estimate what has been neglected when 
discarding the cosine in Eq. \Eqref{sineeq}, 
we compute the residual fluctuations of 
$\hat C_{\bf k}$ and the cross correlation $\hat C_{\bf k} \hat S_{\bf k}$.
To leading order in $n_k$, one has
\ba \label{ABvariances}
 \langle \hat S_{\bf k} \hat S_{{\bf k}'}^{\dagger} \rangle_{in} &=& 
 \left( n_k + \frac{1}{2} - Re(c_k) \right) \, \delta^{3}({\bf k} - {\bf k}') 
 = 2n_k  \, \delta^{3}({\bf k} - {\bf k}')  
 \, , \nonumber \\  
 \langle \hat C_{\bf k} \hat C_{{\bf k}'}^{\dagger} \rangle_{in} &=& 
 \left( n_k + \frac{1}{2} + Re(c_k) \right) \, \delta^{3}({\bf k} - {\bf k}')  
  = O(n_k^{1/4}) \, \delta^{3}({\bf k} - {\bf k}')  
 \, , \nonumber \\
 \langle \hat C_{\bf k} \hat S_{{\bf k}'}^\dagger \rangle_{in} &=& 
 - \left(Im(c_k) + \frac{i}{2}\right) \, \delta^{3}({\bf k} - {\bf k}') 
 = O(n_k^{1/4}) \, \delta^{3}({\bf k} - {\bf k}') 
  \, . 
\ea
When divided by the variance
of $\hat S_{\bf k}$, the variance of $\hat C_{\bf k}$ and 
the cross-correlation are of order $n_k^{-3/4} \sim (k\eta_r)^3
\sim 10^{-75}$. 
One can therefore safely use the distribution Eq. \Eqref{Peff} 
in replacement of the quantum distribution 
Eqs. \Eqref{inoutvac} or \Eqref{ABvariances} 
when calculating the power spectrum.
However, 
in general, this is {\it not} the case for the entropy.

\subsection{Drawbacks of the simplified description}

The simplified description in terms of a statistical ensemble
of sine standing waves has indeed several shortcomings. 
It is of value to describe them with some attention.

First, it should be pointed out that in the early universe, different physical
processes could give rise to different final density matrices, and hence, to
different entropies. 
As we shall see bellow, only a narrow set of these quantum distributions can
be put in correspondence with that of Eq. \Eqref{Peff}.
To describe the 
most general isotropic and Gaussian distribution,
it is necessary to return to two-mode distributions
characterized by 
$n_k$ and $c_k$ of Eq. \Eqref{momenta}.
Second, 
the classical ensemble of sine waves should be considered only
as an effective description of the density matrix.
The appropriate basis to describe this matrix is provided 
by coherent states. The reasons are the following.
On the one hand, they constitute the quantum counterparts of 
classical configurations in 
phase space. Therefore they provide an adequate basis for studying 
the semi-classical limit. In particular, since the spread of coherent states
have no preferred direction in phase space, it will 
be easy to see whether or not a distribution has kept some squeezing.
On the other hand, they are the preferred basis in
which squeezed states decohere when weak interactions are taken into 
account\cite{Zurek93}. 

\section{Two-mode coherent states}

When using coherent states in cosmology, one must pay attention to 
the entanglement between $\bf k$ and $-{\bf k}$ modes. 
A naive use of coherent states which would assign 
amplitudes to each mode separately could erase these
correlations and therefore suppress the information about the temporal phase
of the modes. 
Taking into account the 
entanglement leads to the notion of 'two-mode
coherent states'.

\subsection{Representation of the in vacuum with two-mode coherent states}

To understand the usefulness of two-mode coherent states
it is appropriate to first mention the following properties \cite{CP04}.
Consider a mode ${\bf k}$ in a coherent state 
$\ket{v,\, {\bf k}}$ constructed with out operators, see Appendix A for its
definition. 
Then compute the one-mode reduced state obtained by
projecting it on the two-mode initial vacuum of Eq. \Eqref{in2mode}
\ba
\label{remar}
 \langle v,\, {\bf k} \kett{0 in,\, {\bf k}}  = 
 {\cal A}_k(v) \, \ket{z_k v, \, -{\bf k} } \, .
\ea 
It is remarkable that the state of the $-{\bf k}$ mode
which is entangled to $\ket{v,\, {\bf k}}$ 
is also a coherent state. Its amplitude 
is given by 
the complex conjugate of $v$ 
times $z_k$ characterizing the pair creation process.
These facts follow from the 
EPR correlations in the initial vacuum displayed in Eq. \Eqref{inoutvac}. 
The prefactor ${\cal{A}}_k(v)$ is
\ba \label{A(v)}
 {\cal{A}}_k(v) &=& \inv{\abs{\alpha_k}} 
 \exp{\left(-\frac{\abs{v}^2}{2\abs{\alpha_k}^2} \right)}  \, .
\ea
It is the probability amplitude to find the mode ${\bf k}$ in 
the coherent state $\ket{v, \, {\bf k}}$ given that we start with vacuum at the
onset of inflation.

Using the representation of the identity with coherent states,
Eq. \Eqref{coh1}, 
the two-mode in vacuum can be thus 
decomposed as a {\it single} sum of {\it two-mode coherent states}
 rather than two independent integrations over one-mode coherent
states. In fact we have
\ba \label{insuperpcoh}
 \kett{0 in,\, {\bf k}}  = 
 \int\!\!\frac{d^2v}{\pi} \, 
 {\cal{A}}_k(v) \, \,  \ket{v,\, {\bf k}}  \otimes \ket{z_k v^*, \, -{\bf k}} 
  \, .
\ea
This 
result is exact\footnote{Notice however that the above decomposition is not unique
since the coherent states are not orthogonal, compare with \cite{Kim00}. 
Eq. \Eqref{insuperpcoh} has the advantage to be directly related to
the detection 
of a quasi-classical configuration in the ${\bf k}$ sector.}
and applies even for low occupation numbers.
It is the consequence of the coherence of
the in vacuum and 
holds for every homogeneous pair creation process.

Another important consequence of Eq. \Eqref{remar} is that
the probability to find {\it simultaneously} the ${\bf k}$-mode with
coherent amplitude $v$ and its partner with amplitude $w$ is 
\ba \label{vwdistrib}
 {\cal P}_{2, \, k}(v,w) = 
 \abs{\bra{v,\, {\bf k}}\bra{w,\, -{\bf k}} 0 in, \, {\bf k} \rangle_2}^2 = 
 \abs{{\cal{A}}_k(v)}^2 \times e^{-\abs{w-z_k v^{*}}^2} \, .
\ea 
The second factor arises follows from the overlap between 
two different coherent states: 
$\abs{\langle u \vert v \rangle}^2 = \exp({-\abs{u-v}^2}) $.
Equation \Eqref{vwdistrib} implies that 
once the amplitude of the ${\bf k}$-mode has been measured, the
conditional probability to find its partner in a coherent state $\ket{w}$ is
centered
around $w=z_k v^{*}$. In the high occupation number limit we
are dealing with, the spread ($=1$)
around this mean value is negligible when compared
to the spread in $v$ which is given by $\abs{\alpha_k}^2 = n_k + 1$. 
Therefore, when computing
expectation values in leading order in $n_k$, the
conditional probability acts as a delta function on both 
the real and the imaginary part of $w$. 
This is how the EPR correlations in the in-vacuum 
translate in the coherent states basis.
This 
result 
determines the properties 
%
of the local correlations in the primordial spectra\cite{CP04}.

To complete this analysis, and in preparation for studying decoherence, 
it is also interesting to explicitly write the non-diagonal matrix
elements of the in vacuum density matrix. One has
\ba \label{vwrhov'w'}
 \bra{v}\bra{w} \hat \rho_{in} \ket{v'} \ket{w'} &=& 
 {\cal A}_{2, \, k}(v,w) \, {\cal A}_{2, \, k}(v',w')^{*} \, , 
\ea
where the two-mode amplitude is
\ba
 {\cal A}_{2, \, k}(v,w) &=& 
{\cal A}_k(v) \,
 e^{-\frac{1}{2}\abs{w-z_k v^*}^2 }\,
 e^{i\, {\rm Im}(w^*z_k v^*)} \, .
\ea
Since the initial vacuum is a pure state and since 
$\abs{\alpha_k}^2 \gg 1$,
the above matrix elements do not
vanish, {\it even for macroscopically different coherent states}.
Therefore this distribution does not describe a classical ensemble of these
quasi-classical states.
Fortunately, 
such quantum distributions are 
unstable to any weak perturbation in that they rapidly evolve 
into statistical mixtures.
Let us now describe this decoherence process.


\subsection{Zurek {\it et al.} analysis and minimal decoherence scheme}

In general, it is a difficult question to determine into 
what mixture an initial density matrix will evolve when taking into account
some interactions amongst modes or with other modes.
There exist  however several cases where clear conclusions
can be drawn. First,
when one can neglect the free Hamiltonian, 
the preferred states (that is the basis into which
the reduced density matrix will become diagonal) 
are the eigenstates
of the interaction Hamiltonian \cite{gottfried,Zurek81,Zurek82}.
This approach has been applied in \cite{KiefPolStar98a}, 
to primordial density fluctuations when the (physical) modes are 
almost constant because their wave length is much larger than the
Hubble radius. The conclusion 
is that the preferred basis is provided by amplitude (position) 
eigenstates.
However this conclusion leaves some ambiguity and
might lead to some difficulties.
First, position eigenstates are not normalized.
Second, and more importantly, the spread in momentum is infinite
for these states. Therefore, the velocity field would not
be well defined when the modes re-enter
the horizon. Moreover, as pointed out in \cite{KiefPolStar00},
some additional decoherence could be obtained as they re-enter
the horizon.
In this case, the momentum should be treated 
in the same footing as the position. To cure these problems, 
 some finite spread in position should be introduced.
One then needs a physical criterion to choose this spread.

To remove this ambiguity
it is appropriate to appeal to coherent states
both for mathematical and physical reasons. 
In this article, we shall only present the basic mathematical 
results.
We reserve for a forthcoming publication the justification of 
the physical relevance of these states in a cosmological context. 
Let us simply notice the following points. 
In inflationary cosmology, modes are
weakly interacting harmonic oscillators\cite{Maldacena03}. 
Indeed, given that the relative density fluctuations
have small amplitude ($\sim 10^{-5}$), the hypothesis
of weak interactions is perfectly legitimate.
Second, coherent states
provide the basis in which the density matrix decoheres 
when considering weakly interacting harmonic oscillators.
This  has been shown by 
Zurek \& {\it al.} \cite{Zurek93}. 
The criterion they used to reach this conclusion is
the minimization of the growth of entropy in the course of the evolution. 
With this criterion,
coherent states are more stable than
squeezed states in that the growth of entropy 
one obtains when they are used as initial states is much slower.
Hence, 
when starting with a squeezed state, 
there is a phase of rapid growth of the entropy 
which sends the system into a mixture of coherent states
and which is followed by a period of slower increase. 
The entropy growth is in fact directly related to
the decay of the squeezing. 
Since it is unlikely that the interactions
in the early cosmology could be sufficiently weak so as
to keep some squeezing, we can use the following mathematical result
to infer that the actual entropy of the final distribution
should be higher than (or equal to) a certain bound.

Coherent states indeed define a minimal decoherence scheme 
in the following sense. 
Consider the set of final distributions which result from the
initial distribution of Eq. \Eqref{vwrhov'w'}
through some decoherence process
and which no longer possess any squeezed direction.  
The lowest value of the entropy in this set
is given by the entropy of the incoherent 
superposition of coherent states,
with statistical weights given
by the probabilities to find the corresponding coherent states,
as in Eq. \Eqref{vwdistrib}. 
This distribution gives a lowest entropy simply because 
coherent states have minimal constant spreads (given quantum uncertainties
and when considering free evolution, see Appendix A).
We shall now explicitely write down this distribution 
and compute the entropy it carries.

\subsection{Application to cosmology}
 
When considering the initial distribution 
Eq. \Eqref{vwrhov'w'}, one obtains the following 
density matrix:
\ba \label{rhodec}
 \hat \rho_{min} = \int\!\!\frac{d^2v}{\pi} \frac{d^2w}{\pi}\, 
 {\cal{P}}_{2, \, k}(v,w) \, \,
 \ket{v,\, {\bf k}}\bra{v,\, {\bf k}} \otimes 
 \ket{w, \, -{\bf k} }\bra{w, \, -{\bf k}} \, ,
\ea
where the probability distribution is given in Eq. \Eqref{vwdistrib}. 
In Appendix B we show that Eq. \Eqref{rhodec} is indeed 
the resulting normalized distribution. The technical point 
which requires clarification is the extension 
of the above mentionned minimal scheme
to two-mode squeezed states.

It should first be noted that when computing expectation
values in leading order in $n_k$, the above distribution can be simplified and
written as a single sum of two-mode coherent states,
as in eq. \Eqref{insuperpcoh}:
\ba \label{rhodec2}
 \hat \rho_{min} \simeq \int\!\!\frac{d^2v}{\pi} \,
 \abs{{\cal A}_k(v)}^2  \,\,
 \ket{v,\, {\bf k}} \bra{v,\, {\bf k}} 
 \otimes
 \ket{z v^*, \, -{\bf k} }
 \bra{z v^*, \, -{\bf k}} \, .
\ea
As we shall progressively see, this distribution should be conceived
as the quantum counterpart of the effective distribution of sine
functions discussed in section II.B.

Secondly by reducing the density matrix, some
entropy has been introduced. One verifies indeed that
 ${\rm Tr}(\hat \rho_{min}^2) < 1$.
The important point is 
that this decohered distribution has retained all the information about 
the temporal coherence of the modes. 
In fact one has
\ba \label{RLcoherence}
 {\rm Tr}(\hat \rho_{min} \hat a_{\bf k} \hat a_{-{\bf k}} ) &=&
 \int\!\!\frac{d^2v}{\pi} \frac{d^2w}{\pi}
 \left[ {\cal{P}}_{2, \, k}(v,w) \, v\, w \right] 
 =\int\!\!\frac{d^2v}{\pi} \left[ \abs{{\cal A}_k(v)}^2 v\, (z_kv^*)\right]  = 
 z_k \abs{\alpha_k}^2 \,  ,
 \nonumber\\
 {\rm Tr}(\hat \rho_{min} \hat a_{\bf k}^{\dagger} \hat a_{\bf k} ) &=& 
 {\rm Tr}(\hat \rho_{min} \hat a_{-{\bf k}}^{\dagger} \hat a_{-{\bf k}} ) =
 \abs{\alpha_k}^2 = n_k + 1 \, .
\ea
The first line shows that the cross term is equal 
to that of the original distribution,
see Eqs. (\ref{momenta}, \ref{cterm}). 
In the second line, one sees that
the occupation numbers slightly differ, but to order $1/n_k$ only.

The effect  of having increased by $1$ the occupation number while having kept
untouched the cross-correlation means two things. 
First, the (relative) degree of coherence
has been reduced and therefore
some entropy has been created.
Secondly, in the high occupation limit, the two-point function of
Eq. \Eqref{Ginx} is 
not affected by this loss of coherence: for relevant modes,
 the relative change being of the order of $1/n_k \sim 10^{-100}$.


\subsection{Minimal entropy and the neglect of the decaying mode}

The entropy of any Gaussian two-mode distribution can be exactly
calculated \cite{Zeh85,Serafini04} by using the fact that  
the density matrix of a two-mode squeezed state 
%
is unitarily equivalent to the tensorial product of 
two thermal density matrices of oscillators, see Appendix C.
We shall name $a$ and $b$ these two real oscillators.
One has
\ba \label{map}
 \hat \rho_{2,\, k} = {\cal M}^{\dagger} \, \hat \rho_{th,\, a} \otimes 
 \hat \rho_{th,\, b} \, {\cal M} \, ,
\ea
where $\cal M$ is a unitary operator acting on the two-mode Hilbert space. 
The expression of the (von Neumann) entropy immediately follows:
\ba
 S\left[ \hat \rho_{2,\, k} \right] &=& S\left[ \hat \rho_{th,\, a} \right] +
 S\left[ \hat \rho_{th,\, b} \right] \, ,
\ea
where the entropy of a thermal bath with mean occupation $\bar n$ is
\ba
 S\left[ \hat \rho_{th} \right] = (\bar n +1 ) \ln(\bar n +1 ) - 
 \bar n \ln(\bar n) \, .
\ea
When considering only distributions preserving homogeneity and isotropy,
the occupation numbers of the thermal matrices are equal and
given by
\ba \label{tilde n}
 \bar n_{k} + \frac{1}{2} = 
 \left( (n_{k} + \frac{1}{2})^2 - \abs{c_{k}}^2 \right)^{1/2} \, ,
\ea
where $n_k$ and $c_k$ are defined in Eq. \Eqref{momenta}.

Let us apply this result to several cases.
First, for the two-mode in vacuum of Eq. \Eqref{inoutvac},
the occupation number and the coherence term are 
related by Eq. \Eqref{cterm}, one has
$\bar n_k =0$ as expected. Hence 
the entropy vanishes.

For the decohered matrix Eq. \Eqref{rhodec}, 
using Eq. (\ref{RLcoherence}), 
the occupation number of the two thermal baths are
\ba  
 \bar n_{k} = 
 \frac{1}{2} \left( -1+\sqrt{8(n_k +1 ) +1}\right) \sim
 \sqrt{2  n_k} \, , 
\ea
where the last term is the leading order when 
$n_k \gg 1$.
The two-mode entropy of this mixture is then
\ba
\label{Smin}
 S\left[ \hat \rho_{min} \right] &=& 2 \, S\left[ \hat \rho_{th}
 \right] = 2 \ln \bar n_k + O(1) \, , 
\nonumber \\
 &=& \ln n_k + O(1) = 2\, r_k +  O(1) \, , 
 \nonumber \\
 &\simeq& 100 \ln(10)  \, .
\ea
In the second line, we have expressed the occupation number in term of the
squeezing parameter $r_k$ : $n_k = \sh^2 r_k$. 
Hence, a two-mode squeezed vacuum
state which decoheres in the two-mode coherent basis goes along with an entropy
of $S_{{\bf k},\, -{\bf k}} = 2 \, r_k$ {\it per two-mode}. 
This value is large, but not maximal. Indeed, 
had the coherence term $c_k$ vanished while keeping the occupation numbers 
fixed\cite{MukhaBrandyProkopecg}, 
one would have found the maximal value of the entropy 
which is given by twice this above value, i.e.
\ba
S_{{\bf k},\, -{\bf k}}^{max}= S_{{\bf k},\, -{\bf k}}^{inc} = 4 \, r_k\, ,
\ea 
or $S^{inc}_{\bf k} = 2 \, r_k$ per mode ${\bf k}$.

It is interesting to notice that the entropy
associated with the Gaussian distribution Eq. \Eqref{Peff} 
of sine functions
equals that of Eq. \Eqref{rhodec}, up to an arbitrary constant
which arises from the usual 
ambiguity of attributing an entropy to a classical distribution. 
(This ambiguity can be lifted when introducing $\hbar$ to normalize
the phase space integral.) Using this trick, the entropy associated
with 
Eq. \Eqref{Peff} is $S_{e\!f\!f}= 2 \, r_k$ for each 
{\it independent} mode, since this entropy is maximal. 
However, for each 
{independent} mode here
means for each two-mode since the mode $-{\bf k }$ 
is no longer independent of the ${\bf k }$ mode once
the cosines have been neglected, see Eq. \Eqref{Aeff}.
From this equality of entropies, we conclude that the
quantum density matrix which corresponds to the Gaussian
ensemble of sine functions is precisely given by Eq. \Eqref{rhodec}.

A priory one might think that many quantum distributions 
can be associated with the classical distribution Eq. \Eqref{Peff}.  
This is not the case when imposing that Gaussianity is preserved
and that the entropies coincide. Indeed, 
in the high squeezing limit, 
the entropy is an
extremely sensitive function of the relative coherence. 
To see this dependence,
let us calculate the entropy of a generic distribution \Eqref{momenta},  
and let us write the norm of the coherence term as
\ba
\label{paramd}
 \abs{c_k}^2 = n_k(n_k + 1 - \delta_k) \, ,
\ea
where $\delta_k$ is a real number between $0$ and $n_k + 1$.
The equation \Eqref{tilde n} has the solution
$\bar n(\bar n +1 ) =  n_k \delta_k $. 
They are three characteristic values. 
$\delta_k =0$ obviously corresponds to the pure squeezed state: the in vacuum.
$\delta_k = 1$ corresponds to the minimal decoherence scheme
with entropy given in Eq. \Eqref{Smin}. $\delta_k = n_k + 1$
corresponds to the thermal case with maximum entropy. 
From this analysis we see that 
the uncertainty in the definition of the quantum distributions
which give rise to the entropy $S_{{\bf k},\, -{\bf k}}= 2 r_k$ is 
very limited: $\delta_k$ must be of order $1$ in the following sense.
Consider that the loss of coherence scales as
$\delta_k \propto n_k^\gamma$. Then the thermal occupation number and the entropy
respectively scale as $\bar n_k \propto n_k^{(1+\gamma)/2}$ and
$S_{{\bf k},\, -{\bf k}} = (1+\gamma) 2 r_k + Const$. 
This linear dependence in $r_k$
implies that the distributions with entropy given
by Eq. \Eqref{Smin} all have $\gamma =0$.

We re-emphasize that a value of $\delta_k$ smaller than $1$ is 
unlikely in the context of
primordial fluctuations since it would mean that the distribution
has kept some of its quantum squeezeness. The remaining question thus
concerns the computation of $\delta_k$, noticing that it can receive
contributions both from the inflationary period and from the adiabatic 
era\cite{KiefPolStar00}.
The challenge is to determine which one is more important and
what could be a realistic value of $\delta_k$.

As a final comment, we provide an upper bound for the 
decoherence entropy which could have resulted from processes
in the inflationary phase. Because increasing the decoherence implies
increasing the power of the growing mode,
one obtains a bound on the decoherence level 
when requiring that the power of the decaying mode
be equal to that of the growing mode at the onset of the adiabatic era.
(This requirement follows from the fact
that the rms value of the primordial 
fluctuations (of the Bardeen potential) cannot be 
much higher than that obtained from in vacuum
because otherwise this would invalidate the whole
framework of linear metric perturbations.)
Using Eq. \Eqref{Ginx} evaluated at $\eta_r$
and the parameterization of Eq. \Eqref{paramd}, one obtains
\ba
\delta_k = n_k^{1/2}\, , \qquad S^{upper}_{{\bf k},\, -{\bf k}}= 3 r_k + O(1)\ .
\ea
If no further decoherence is added in the adiabatic
area, this should be the maximum amount of entropy stored
in the primordial spectrum. Notice that when
evaluated at recombination, the two-point
function is still unaffected by this modification of the coherence
because at that time the decaying mode has still further decreased.
Indeed the residual modifications are then of the order of $n_k^{-1/2}
\sim 10^{-50}$.


\vskip .5 truecm

{\bf Acknowledgments:} 
We are grateful to Dani Arteaga, 
Claus Kiefer, Serge Massar, 
Jihad Mourad and Alexei Starobinsky
for useful remarks.

\begin{appendix}
\section{Coherent states}

This appendix aims to present 
the properties which we shall use
in the body of the manuscript. 
For more details, we refer to \cite{Glauber63a,Glauber63b,Zhang90}.

Coherent states (of a real oscillator) can be 
defined as eigenstates of the annihilation operator:
\ba \label{DefCoh1}
 \hat a \ket{v} = v \ket{v} \, ,
\ea
where $v$ is a complex number.
In Fock basis it is written as
\ba \label{DefCoh2}
 \ket{v} = e^{-\frac{\vert v \vert^2}{2}}  \sum_{n=0}^{\infty}
 \frac{v^n}{\sqrt{n!}} \ket{n} \, ,
\ea  
where the exponential prefactor guarantees that the state is normalized 
to unity $\langle v \vert v \rangle=1$.
They are also obtained by the action of a displacement operator on the vacuum :
\ba \label{displacement}
 \ket{v} = \hat D(v) \ket{0} = e^{v^* \hat a - v \hat a^{\dagger}} \, .
\ea

The first interesting property of coherent states is that they correspond 
to states with a well defined complex amplitude $v$. 
Indeed, by definition \Eqref{DefCoh1}, 
the expectation values of the annihilation and creation operators are
\ba
 \bra{v} \hat a \ket{v} = v \, , \quad \bra{v} \hat a^{\dagger} \ket{v} = v^{*} 
 \, .
\ea
Thus the mean occupation number is
\ba \label{occupnumber}
 \bra{v} \hat a^{\dagger} \hat a \ket{v} = \abs{v}^2 \, .
\ea 
It is to be also stressed that the variances vanish:
\ba \label{nullvariance}
 \Delta \hat a^2 = \bra{v} \hat a^2 \ket{v} - \bra{v} \hat a \ket{v}^2 = 0 \, , 
 \quad \Delta \hat a^{\dagger \, 2} = \bra{v} \hat a^{\dagger \, 2} \ket{v} - 
 \bra{v} \hat a^{\dagger} \ket{v}^2 =  0 \, .
\ea 

From these properties one sees that the expectation values
of the position and momentum operators (in the Heisenberg picture) 
\ba 
 \hat q(t)=\sqrt{\frac{\hbar}{2\omega}}\left(\hat a e^{-i \omega t} + 
 		\hat a^{\dagger} e^{i \omega t}\right) \, , \qquad 
 \hat p(t)=-i\sqrt\frac{\hbar\omega}{2}(\hat a e^{-i \omega t} - 
 				\hat a^{\dagger}e^{i \omega t}) 
 \, , \nonumber
\ea
are
\ba \label{cltrajincohstate}
 \bar q(t) &=& \bra{v} \hat q(t) \ket{v} = 
 \sqrt{\frac{\hbar}{2\omega}}( v e^{-i \omega t} + v^* e^{i \omega t}) 
            = \sqrt{\frac{2\hbar}{\omega}} \abs{v} \cos(\omega t - \phi_v) 
 \, , \nonumber \\ 
 \bar p(t) &=& \bra{v} \hat p(t) \ket{v} = 
  -i\sqrt{\frac{\hbar\omega}{2}} ( v e^{-i \omega t} - v^* e^{i \omega t}) 
            = -\sqrt{2\hbar\omega} \abs{v} \sin(\omega t - \phi_v) 
	    = \partial_t \bar q(t) \, .
\ea
We have used the polar decomposition $v=\abs{v} e^{i \phi_v}$.
These expectation values have a well defined amplitude and phase and 
follow a classical trajectory of the oscillator. 
This is due to the ``stability'' of coherent states under the
evolution of the free Hamiltonian $2 H_0= p^2 + \omega^2 q^2$:
if the state is $\ket{v}$ at time $t_0$, one immediately gets from 
\Eqref{DefCoh2} that at a later time $t$, the state is 
given by $\ket{v(t)} = \ket{ve^{-i\omega (t-t_0)}}$.
Notice that the variances of the position and the momentum are 
\ba
 \Delta \hat q^2 = \frac{\hbar}{2\omega} \, , \qquad  
 \Delta \hat p^2  = \frac{\hbar\omega}{2} \, .
\ea
They minimize the Heisenberg uncertainty relations and are time-independent.
Hence, in the phase space $(q,p)$, a coherent state can be
considered as a unit quantum cell $2\pi \hbar$ in physical units 
(see also \Eqref{measure} for the measure of
integration over phase space) centered on the classical position and momentum of
the harmonic oscillator $(\bar q(t),\bar p(t))$. 
In the large occupation number limit $\abs{v}\gg 1$,
coherent states can therefore be
interpreted as classical states since 
$\Delta \hat q / \sqrt{\bar q^2+ \bar p^2/\omega^2} =
\Delta \hat p / \sqrt{\omega^2\bar q^2+ \bar p^2} = 1/2\abs{v}$.
This is a special application of the fact that coherent states
can in general be used to define the classical limit 
of a quantum theory, see \cite{Zhang90} and references therein.

One advantage of coherent  states \cite{Glauber63a} is that the 
calculations of Green functions resembles closely to those of the 
corresponding classical theory (i.e. treating the fields not as operators 
but as c-numbers) provided either one uses normal ordering,
or one considers only the dominant contribution when $\abs{v}\gg 1$.
We compute the Wightman function in the coherent state $\ket{v}$
\ba
\widetilde G_v(t,t') &=& \bra{v} \hat q(t) \hat q(t') \ket{v} \nonumber \\
           &=& \langle :\hat q(t)\hat q(t') : \rangle_v + 
	   \frac{\hbar}{2\omega} e^{i\omega(t-t')} \, ,
\ea
where we have isolated the contribution of the vacuum.
The normal ordered correlator is order $\abs{v}^2$: 
\ba \label{cohgreen}
 \langle :\hat q(t)\hat q(t') : \rangle_v &=& 
 \frac{\hbar}{\omega} \, 
 Re\left[\langle \hat a^2 \rangle_v e^{-i\omega(t+t')} + 
    \langle  \hat a^{\dagger} \hat a \rangle_v e^{i\omega(t-t')} \right]
    \nonumber  \\ 
 &=& \frac{2\hbar}{\omega} \vert v \vert^2 \cos(\omega t - \phi_v) 
     \cos(\omega t' -\phi_v)   = \bar q(t) \,  \bar q(t') \, .
\ea
We see that the perfect coherence of the state, namely 
$\abs{\langle \hat a  \hat a  \rangle_v } 
= \langle \hat a^\dagger  \hat a  \rangle_v  $ 
is necessary to 
combine the contributions of the diagonal and the interfering term so as to
bring the time-dependent classical position $ \bar q(t)$ in 
Eq. \Eqref{cohgreen}.

The wave-function of a coherent state in the coordinate
representation is given by 
\ba \label{CohWaveFunct}
  \psi_{v}(q) = \left(\frac{\omega}{\pi\hbar}\right)^{1/4} 
 \exp\left(-\frac{\omega}{2\hbar}(q-\bar q)^2 -i \frac{\bar p q}{\hbar} \right) 
 \, ,
\ea
where $v=(\omega \bar q + i \bar p)/\sqrt{2\omega\hbar}$.
This follows from the definition 
$\bra{q} \hat a \ket{v} = v \langle q \vert v \rangle$. 
From this equation one notes that two coherent states are not orthogonal.
The overlap between two coherent states is
\ba \label{overlap}
 \langle v \vert w \rangle = \exp\left(v^{*} w - \inv{2}\abs{v}^2 - 
 \inv{2}\abs{w}^2 \right)  \, .
\ea
Nevertheless they form an (over)complete
basis of the Hilbert space in that 
the identity operator in the
coherent state representation $\{ \ket{v} \}$ reads
\ba
\label{coh1}
 {\bf 1} = \int\!\!\frac{d^2v}{\pi} \, \ket{v} \bra{v} \, .
\ea 
The measure is 
\ba \label{measure}
 \frac{d^2v}{\pi} = \frac{d({\mathrm Re} v) d({\mathrm Im} v)}{\pi} = 
 \frac{d\bar q d\bar p}{2\pi \hbar} \, .
\ea
The representation of identity can be established by calculating 
the matrix elements of both sides of
the equality in the coordinate representation $\{ \ket{q} \}$, 
with the help of \Eqref{CohWaveFunct}. 


\section{Application of Zurek \& al. results to the cosmological case}

In this appendix we show that Eq. \Eqref{vwrhov'w'}
is indeed the minimal decohered distribution by
decomposing the 
complex 
mode $\hat \phi_{\bf k}$ into
two real oscillators $\hat \phi_{1}$ and $\hat \phi_{2}$ given by
its real and imaginary parts.
Since the two-mode Hamiltonian is Hermitian, it splits into the sum 
of two identical one-mode oscillator Hamiltonians 
for $\hat \phi_{1}$ and $\hat \phi_{2}$ separately. 
Notice that the annihilation operators of these two real oscillators,
$\hat a_1 = \left(\hat a_{\bf k} + \hat a_{-\bf k}\right)/\sqrt{2} , \, 
\hat a_2 = -i\, \left(\hat a_{\bf k} - \hat a_{-\bf k}\right)/\sqrt{2}$, 
mix ${\bf k}$ and $-{\bf k}$ annihilation operators. 
Hence they can easily take
into account the entanglement between these two modes.    

A two-mode squeezed state $\kett{0 in, {\bf k}}$ 
can always be written as the tensorial
product of the two one-mode squeezed states\cite{schum}.
In our case, the one-mode squeezed states
are those of the oscillators $1$ and $2$ because the
Hamiltonian separates. Thus we have
\ba
 \kett{0 in, {\bf k}} = \ket{0 in, 1} \otimes \ket{0 in, 2} \, . 
\ea
The one-mode squeezed states are governed by the same parameter $z/2$ :
\ba
 \ket{0 in, 1} = \frac{1}{\sqrt{\abs{\alpha}}} \sum\limits_{n=0}^{\infty} 
 \left( \frac{z}{2} \right)^n \frac{\sqrt{2n !}}{n!} \ket{2 n,\, 1} \, .
\ea
The same expression holds for the ket  $\ket{0 in, 2}$.

The overlap of this one-mode squeezed state with a one-mode coherent state is
\ba
 \langle v ,\, 1 \vert 0 in, 1  \rangle = 
 \frac{1}{\sqrt{\abs{\alpha}}} \exp\left(- \abs{v_1}^2/2+ zv_1^{*\, 2}/2 \right) \, .
\ea

According to \cite{Zurek93}, when taking small
interactions into account, the density matrix of a
one-mode squeezed state will preferable decohere into the mixture
\ba
 \hat \rho_{red,\, 1} =  \int\!\!\frac{d^2 v_1}{\pi} \, 
  {\cal P}_1(v_1)  \, 
 \ket{v_1} \bra{v_1} \, .
\ea
where the statistical weight is given by the probability to find a coherent
state starting from the in vacuum :
\ba \label{proba1}
 {\cal P}_1(v_1) = \abs{\langle v ,\, 1 \vert 0 in, 1  \rangle }^2 &=&
 \frac{1}{\abs{\alpha}} \exp\left(- \abs{v_1}^2 + Re(zv_1^{*\, 2}) \right) \, ,
 \nonumber \\
 &=& \frac{1}{\abs{\alpha}}e^{- 2 R_1^2} \,
 e^{-  I_1^2/2 \abs{\alpha}^2 + O(n^{-3/4}R_1 I_1)} \, .
\ea
In the second line we have introduced the real and imaginary parts
of $v_1$ in order to show that one gets an ellipse of great axis
equal to $\abs{\alpha}^2$ which is oriented along the imaginary axis.
The width of the small axis is $1/2$, as in vacuum. 


For this decoherence procedure to be valid,
as noticed in \cite{KiefPolStar98a}, it is important that the 
interactions do not break the
coherence between ${\bf k}$ and $-{\bf k}$ modes, or equivalently do not mix 
$\phi_1$ and $\phi_2$. 

The product of two one-mode coherent states $1$ and $2$ is also the
product of a coherent state for the ${\bf k}$ and $-{\bf k}$ modes:
\ba
 \ket{v_1}\otimes\ket{v_2} &=&  
 \hat D_1(v_1) \hat D_2(v_2) \ket{0 in, 1} \otimes \ket{0 in, 2} 
 \, , \nonumber \\
 &=& \hat D_{\bf k}(v) \hat D_{-{\bf k}}(w) 
 \ket{0 in, {\bf k}} \otimes \ket{0 in, -{\bf k}} 
 \, , \nonumber \\
 &=&  \ket{v, {\bf k}} \otimes \ket{w, -{\bf k}} \, ,
\ea 
where the amplitudes are related by 
\ba
 v=\frac{v_1+iv_2}{\sqrt{2}} \, , \qquad w=\frac{v_1-iv_2}{\sqrt{2}} \, .
\ea
Finally, the product of the probabilities \Eqref{proba1} give the probability 
\Eqref{vwdistrib}. Performing the change of variables from $(v_1,\, v_2)$ to 
$(v,\, w)$ completes the proof.



\section{Diagonalization of the covariance matrix}


Introducing the position and momentum variables for each mode, i.e.
$\hat a_{{\bf k}} = (\hat q_{{\bf k}} + i\hat p_{{\bf k}})/\sqrt{2}$ and 
$\hat a_{-{\bf k}} = (\hat q_{-{\bf k}} + i\hat p_{-{\bf k}})/\sqrt{2}$, 
and defining the vector 
\ba
 {\bf \zeta}^{\dagger} = \left( \hat q_{{\bf k}} \, \, \hat p_{{\bf k}} \, \, 
 \hat q_{-{\bf k}} \,\,  \hat p_{-{\bf k}} \right) \, ,
\ea
one has the covariance matrix 
\ba
 C = \langle \left[ \, {\bf \zeta}_i , \, {\bf \zeta}_j \right]_+ \rangle = 
 \left( 
 	\begin{array}{c c c c}
	n_{k} + \frac{1}{2} & 0 & c_r & c_i \\
	0 & n_{k} + \frac{1}{2} & c_i & -c_r \\
	c_r & c_i & n_{k} + \frac{1}{2} & 0 \\
	c_i & -c_r & 0 & n_{k} + \frac{1}{2} 
	\end{array}
 \right) \, ,
\ea
where $\left[ \, , \, \right]_+$ is the anticommutator. 
Notice that $\left( n_{k} + \frac{1}{2}\right)^2 - \abs{c_k}^2 > 0$ 
is a necessary condition for the matrix to have positive eigenvalues.

The transformation Eq. \Eqref{map} amounts to diagonalize 
this matrix:
\sba
 C &=& M^{t} T  M \, , \\
 T &=&  \left( \bar n_{k} + \frac{1}{2}\right) {\bf 1}
\sea
The matrix $T$ is the covariance matrix of the two thermal density matrices
$\hat \rho_{th,\, a} \otimes \hat \rho_{th,\, b}$ in Eq. \Eqref{map}. 
The matrix $M$ is the
product of two local transformations  
and one global rotation $R$. The latter brings the covariance matrix $C$
under a $2\times 2$ bloc diagonal form. A product of 
local rotations $R_1(\theta_1) \oplus R_2(\theta_2)$
diagonalize each bloc, and the product of 
local squeezing $S_1(r_1) \oplus S_2(r_2)$ brings the
resulting matrix under the form $T$. Explicitly one has
\ba
 R(\phi) &=& 
 \left( 
 	\begin{array}{c c c c}
	\cos \phi & 0 & -\sin \phi &  0 \\
	0 & \cos \phi & 0 & -\sin \phi \\
	\sin \phi &  0 & \cos \phi & 0  \\
	 0 & \sin \phi& 0 & \cos \phi 
	\end{array}
 \right) \, , \nonumber \\
 R(\theta) &=& 
 \left( 
 	\begin{array}{c c}
	\cos \theta & \sin \theta \\
	-\sin \theta & \cos \theta  
	\end{array}
 \right)  \, ,\qquad
 S(r) = 
 \left( 
 	\begin{array}{c c}
	e^{r/2} & 0 \\
	0 &  e^{-r/2} 
	\end{array}
 \right)  \, ,
\ea
with the rotation angles $\phi = \pi/4$, $\theta_1=\theta_2$ given by 
$\tan 2\theta = -c_i/c_r$, and the
squeezing parameter $r_1=-r_2$ defined by $\th r = - \abs{c_k}/(n_k + 1/2)$. 

The eigenvalue $\bar n_k$ of the thermal matrices
is easily obtained by conservation of the determinant: 
\ba
 {\rm det} \, C &=& \left( (n_{k} + \frac{1}{2})^2 - \abs{c_k}^2 \right)^2
 = \left(\bar n_{k} + \frac{1}{2}\right)^4 
 \, . \nonumber
\ea

\end{appendix}

\end{document}